\documentstyle[prd,aps,psfig]{revtex} 
\newcommand {\ignore}[1]{}
\def\lf{\leaders\hbox to 1em{\hss.\hss}\hfill}
\def\21{$SU(2) \ot U(1)$}
\def\321{$SU(3) \ot SU(2) \ot U(1)$}
\def\ne{\hbox{$\nu_e$ }}

\def\ns{\hbox{$\nu_{s}$ }}

\def\bne{\hbox{$\bar{\nu}_e$ }}

\def\bns{\hbox{$\bar{\nu}_s$ }}

\def\eq#1{{eq. (\ref{#1})}}

\def\fig#1{{Fig. \ref{#1}}}

\def\VEV#1{\left\langle #1\right\rangle}
\let\vev\VEV

\def\lsim{\raise0.3ex\hbox{$\;<$\kern-0.75em\raise-1.1ex
\hbox{$\sim\;$}}}
\def\gsim{\raise0.3ex\hbox{$\;>$\kern-0.75em\raise-1.1ex
\hbox{$\sim\;$}}}

\def\beq{\begin{equation}}
\def\eeq{\end{equation}}
\def\bef{\begin{figure}}
\def\eef{\end{figure}}
\def\bet{\begin{table}}
\def\eet{\end{table}}
\def\bea{\begin{eqnarray}}
\def\eea{\end{eqnarray}}
\def\ba{\begin{array}}
\def\ea{\end{array}}
\def\bi{\begin{itemize}}
\def\ei{\end{itemize}}
\def\ben{\begin{enumerate}}
\def\een{\end{enumerate}}
\def\ot{\otimes}
\def\apj#1#2#3{          { Astrophys. J. }{\bf #1}, #3 (19#2)}

\def\aa#1#2#3{          { Astron. \& Astrophys.  }{\bf #1}, #3 (19#2)}

\def\ib#1#2#3{           {\it ibid. }{\bf #1}, #3 (19#2)}

\def\np#1#2#3{           { Nucl. Phys. }{\bf #1}, #3 (19#2)}
\def\pl#1#2#3{           { Phys. Lett. }{\bf #1}, #3 (19#2)}
\def\pr#1#2#3{           { Phys. Rev. }{\bf #1}, #3 (19#2)}

\def\prl#1#2#3{          { Phys. Rev. Lett. }{\bf #1}, #3 (19#2)}

\def\n.c.#1#2#3{         { Nuovo Cim. }{\bf #1}, #3 (19#2)}
\def\r.n.c.#1#2#3{       { Riv. del Nuovo Cim. }{\bf #1}, #3 (19#2)}
\def\sjnp#1#2#3{         { Sov. J. Nucl. Phys. }{\bf #1}, #3 (19#2)}

\renewcommand{\thefootnote}{\fnsymbol{footnote}}
\def\dfrac#1#2{{\displaystyle\frac{#1}{#2}}}

\draft
\preprint{}

\begin{document}

\title{\vglue -0.8cm  
\hfill \text{astro-ph/9811181}\\
\hfill \text{\small FTUV/98-82, IFIC/98-83}\\
\vglue 0.5cm
Type-II Supernovae and  Neutrino Magnetic Moments}

\author{H. Nunokawa$^1$\thanks{E-mail: nunokawa@ifi.unicamp.br},
R. Tom\`as$^2$  \thanks{E-mail: ricard@flamenco.ific.uv.es} and 
J. W. F. Valle$^2$\thanks{E-mail: valle@flamenco.ific.uv.es}}

\address{
$^1$Instituto de F\'{\i}sica Gleb Wataghin\\
Universidade Estadual de Campinas - UNICAMP\\
13083-970 Campinas SP Brasil \\
$^2$Instituto de F\'{\i}sica Corpuscular - C.S.I.C.\\
Departament de F\'{\i}sica Te\`orica, Universitat de Val\`encia\\
46100 Burjassot, Val\`encia, SPAIN\\
URL http://neutrinos.uv.es}

\date{November, 1998}

\maketitle
\begin{abstract}
The present solar and atmospheric neutrino data together with the LSND
results and the presence of hot dark matter (HDM) suggest the existence 
of a sterile neutrino at the eV scale.  We have reanalysed the effect of
resonant {\sl sterile} neutrino conversions induced by neutrino
magnetic moments in a type-II supernova. 
We analyse the implications of $\nu_e-\nu_s$ and 
$\bar{\nu}_e-\bar{\nu}_s$ ($\nu_s$ denotes sterile neutrino)
conversions for the supernova shock re-heating, 
the detected $\bar\nu_e$ signal from SN1987A and 
the $r$-process nucleosynthesis hypothesis. 
Using reasonable magnetic field profiles we determine the sensitivity 
of these three arguments to the relevant neutrino parameters, i.e. 
the value of the transition magnetic moment and the $\nu_e-\nu_s$ mass
difference $\Delta m^2 \equiv \Delta m^2_{LSND/HDM}$
\end{abstract}

\vskip 0.5cm
%\\ Key words: supernova neutrinos, neutrino magnetic moment, 
%resonant spin-flavour precession, \\r-process, shock-reheating, SN1987A}
\vglue -1.2cm

% 14.60.Pq:   Neutrino mass and mixing
% 14.60.St:   Non-standard-model neutrinos, right-handed neutrinos, etc.
% 13.15+g:    Neutrino interactions

\renewcommand{\thefootnote}{\arabic{footnote}}
\setcounter{footnote}{0}
\section{Introduction}

It was recognized a while ago~\cite{ptv92} that the simplest model to
reconcile the existence of a hot dark matter (HDM) component in the
Universe \cite{cobe2} and possibly incorporating an explanation of the
LSND results \cite{LSND} with the solar \cite{solarexp} and
atmospheric neutrino data \cite{superkatm98} is to invoke a fourth
sterile (electroweak singlet) neutrino at the eV scale. Recently there
have been many papers on this scheme and its variants as well as
phenomenology~\cite{ptvlate}.
The effects of active to sterile neutrino transitions induced by the
mixing on supernova physics have been studied in a number of papers
\cite{sterilemixing,sterilenprv}.

%Here we will neglect the mixing between \ne and \ns ($\nu_s$ denote
%sterile neutrino), but assume that it has a nonzero transition magnetic
% 
Here we also consider the relevance of such active to sterile 
neutrino conversion $\nu_e \to \nu_s$ ($\nu_s$ denote sterile neutrino) 
but induced by a nonzero transition magnetic moment (TMM), 
similar to the transition magnetic moment between two active neutrino 
flavours~\cite{SFP}.  
Such TMM leads to a neutrino spin-flavour precession (SFP) effect 
similar to the active-active case~\cite{SFP}. 
The same way as the SFP effect may take place resonantly in the presence 
of matter \cite{RSFP}, like MSW effect \cite{MSW}, 
so does the active-sterile SFP. 
Here we adopt the general case where \ne and \ns are two independent 
Majorana particles. 
However, the sterile state can be a right-handed component of a Dirac
neutrino. In this case, $\nu_s$ ($\bar{\nu}_s$) should be regarded as
a particle carrying lepton number $-1$ (1) in our notation.

Although expected to be small, e.g. if the neutrino mass is introduced
{\sl a la Dirac} in the same way as the charged fermion masses in the
standard electroweak theory the resulting neutrino magnetic moment is
known to be very small \cite{FS}, $\mu_\nu\sim 3 \times 10^{-19}
(m_\nu/ 1 \mbox{eV}) \mu_B$ where $m_\nu$ is the neutrino mass
and $\mu_B$ denotes the Bohr magneton.
However, several attempts have been made to construct various
mechanisms to induce a large neutrino magnetic moment of order $\sim
10^{-11} \mu_B$ \cite{magmodels}. In fact this is a natural
possibility in the context of radiative models of neutrino mass, such
as that of the sterile neutrino models of refs.~\cite{ptv92,pv93,cm93}.

Let us first recall here the previous bounds on the neutrino magnetic
moments from laboratory experiments as well as from astrophysical
considerations. 
Starting with laboratory limits, the upper bound on the neutrino 
magnetic moments comes from the $\bar{\nu}_e e$ scattering experiments
which give \cite{PDG98}, 
\begin{equation}
\mu_\nu < 1.8 \times 10^{-10} \mu_B. 
\end{equation}
  This bound applies to the
direct or transition magnetic moment of Dirac neutrinos, as well as to
the transition magnetic moment of Majorana neutrinos.

{}From the SN1987A neutrino observations the bounds on dipole magnetic
moment of Dirac neutrinos has been derived by considering the
helicity-flipping scattering processes such as $\nu_L e^- \to \nu_R
e^-$ and $\nu_L p \to \nu_R p$ inside the supernova core
\cite{snbound}.  
By requiring that the $\nu_R$ luminosity should be not too large 
in order to account for the observed neutrino data from SN1987A in 
Kamiokande II \cite{KA} and IMB \cite{IMB} detectors,
the constraint $\mu_\nu \lsim 10^{-12} \mu_B$ is obtained.  
Similar upper bound has been obtained also in a recent paper \cite{Ayala}
by considering neutrino helicity flipping process through photon 
Landau damping in a dense relativistic plasma in a supernova core. 
It has also been discussed \cite{snbound} that this bound could 
be improved to $\mu_\nu \lsim 10^{-13} \mu_B$ by taking into account 
the absence of higher energy neutrino events which would be expected 
due to the spin rotation of higher energy $\nu_R$ produced in the 
core to $\nu_L$ in the galactic magnetic field.  
Let us note that these discussion also applies to the transition magnetic 
moment of Dirac neutrinos as well as to the ones which connects active and 
sterile Majorana neutrinos if one can neglect the mass squared difference 
$\Delta m^2$ in the process discussed above. 
However, it has been pointed out that \cite{voloshin} these bounds 
could be invalid if the resonant re-conversion of neutrinos take place 
in the supernova.

One could also obtain limits from the arguments of excessive cooling
of red giant stars due to the neutrino emission induced by the
neutrino magnetic moment. Ref. \cite{raffelt} gives the bound,
\begin{equation}
\mu_\nu < 3 \times 10^{-12} \mu_B,
\end{equation} 
which also applies both to Dirac magnetic moments, as well as for
Majorana transition moments.

In this paper we show that a non-zero \ne - \ns transition magnetic
moment (TMM) would have an important effects in the presence of
the strong magnetic fields $10^{15}$ Gauss during a SN explosion
\footnote{
It would also have implications for the solar neutrino problem
\cite{akhmedov97,gn98}.} 
since there can be a resonant spin-flavour precession (RSFP) between 
the active and the sterile neutrino flavour, similar to what occurs for
the active-active case. The latter has been extensively considered in
the literature~\cite{activeTM_SN}. We show how \ne - \ns resonant
conversions induced by the active-sterile TMM enhance the $r$-process
nucleosynthesis in all of the mass range relevant for the neutrino hot
dark matter scenario \cite{cobe2}. On the other hand we estimate the
restrictions on neutrino active-sterile TMM from SN1987A data, and
investigate the influence of these \ne - \ns TMM-induced conversions in
the shock revival problem. For previous related work see refs.~
\cite{snbound,voloshin,juha,PSV}. In particular, ref.~\cite{PSV} has
considered \ne - \ns TMM-induced RSFP in the context of SN physics for
the case of random magnetic fields, in which case the conversion
occurs in a non-periodic regime. Here we are considering the case of
regular fields, and we discuss novel issues related to SN
physics which were not considered in \cite{PSV}.
Throughout this paper we assume $\mu_\nu < 10 ^{-12} \mu_B$
so that the neutrino helicity flipping process in the core 
\cite{snbound,Ayala}, discussed above can be neglected 
(see the discussion in Sec. IV for the case if $\mu_\nu > 10 ^{-12}
\mu_B$).

In Sec. II we briefly describe the picture of the neutrino propagation
in matter and the resonant $\nu_e -\nu_s$ conversion \cite{MSW}.
Sec. III.A discusses the implications of $\nu_e -\nu_s$ conversions for
the neutrino re-heating mechanism.  In Sec. III.B we analyse the impact
of our scenario in the later epoch of supernova evolution (few seconds
after the core bounce) for SN (anti-) neutrino detection rates
(Sec. III.B), as well as for $r$-process nucleosynthesis
(Sec. III.C). In Sec. IV we summarize our results.

\section{The active-sterile neutrino resonant spin precession}

In our discussion we only consider the $\nu_e \to \nu_s$ and
$\bar{\nu}_e \to \bar{\nu}_s$ conversion channels, where \ns
($\bar{\nu}_s$) is a sterile neutrino, due to resonant spin
precession. For simplicity we neglect in what follows the mixing
between \ne and \ns so that the $4 \times 4$ evolution Hamiltonian for
the neutrino system \cite{SFP} reduces to an effective $2 \times 2$
system. Moreover we consider the $\Delta m^2 = m^2_s -m^2_e > 0$ case,
i.e., sterile state is heavier. The evolution of the $\nu_e -\nu_s$
system in the matter background with non-zero magnetic field is
determined by the following Schr\"oedinger-like equation,
\begin{equation}
i{\mbox{d} \over \mbox{d}r}\left(\matrix{
\nu_e \cr\ \nu_s\cr }\right) = 
\left(\matrix{
V_e - \frac{\Delta m^2}{2E_\nu} & \mu_\nu B \cr
\mu_\nu B & 0 \cr} \right)
\left(\matrix{
\nu_e \cr\ \nu_s \cr}\right) \,\,, 
\label{evolution1}
\end{equation}
where $\mu_\nu$ is the neutrino magnetic moment and $B$ is the
magnetic field strength perpendicular to the neutrino trajectory.  The
effective potential $V_e$ for $\nu_e$ arises from the coherent
forward neutrino scattering off-matter constituents \cite{MSW} and is
given by \footnote{It is reasonable to neglect the contributions from
the neutrino background to the effective potential, since the neutrino
densities in the relevant regions are small.},
\begin{eqnarray}
\label{potential}
V_e & = &\frac{\sqrt{2}G_F \rho}{m_N} (Y_e- \frac{1}{2}Y_n)=
\frac{\sqrt{2}G_F \rho}{2m_N} (3Y_e- 1)\,, %\,\,\,\,\,\,\,\,\, V_s= 0\,,
\\
Y_e& \equiv & \frac{n_e}{n_e+n_n} \, ,\hskip 1 cm Y_n = 1-Y_e\, .\nonumber
\end{eqnarray}
Note that there is no potential for $\nu_s$, i.e., $V_s =0$.  Here
$G_F$ is the Fermi constant, $\rho$ is the matter density, $m_N$ is
the nucleon mass and $n_e$ and $n_n$ are the net electron and the
neutron number densities in matter, respectively. Note that charge
neutrality $n_p=n_e$ is assumed.  
For the $\bar{\nu}_e \to \bar{\nu}_s$ system the matter potential 
just change its sign.

The resonance condition is:
\begin{equation}
V_e = \frac{\Delta m^2}{2E_\nu}  ~.
\label{rc}
\end{equation}
Let us note that for $\Delta m^2 >0$, either $\nu_e \to \nu_s$ 
(for $V_e > 0$ i.e. $Y_e> 1/3$) or $\bar{\nu}_e \to \bar{\nu}_s$ 
(for $V_e <0$ i.e. $Y_e < 1/3$) conversions take place. 
This is important because,
as we discussed in ref. \cite{sterilenprv}, in the region above the
neutrinosphere the matter potential $V_e$ changes its sign due to the
different chemical content.
%
%%%%%%%%%%%%%%%%%%%%%%%%%%%%%%%%%%%%%%%%%%%%%%%%%%%%%%%%%%%%
\bef[t]
\centerline{\protect\hbox{
\psfig{file=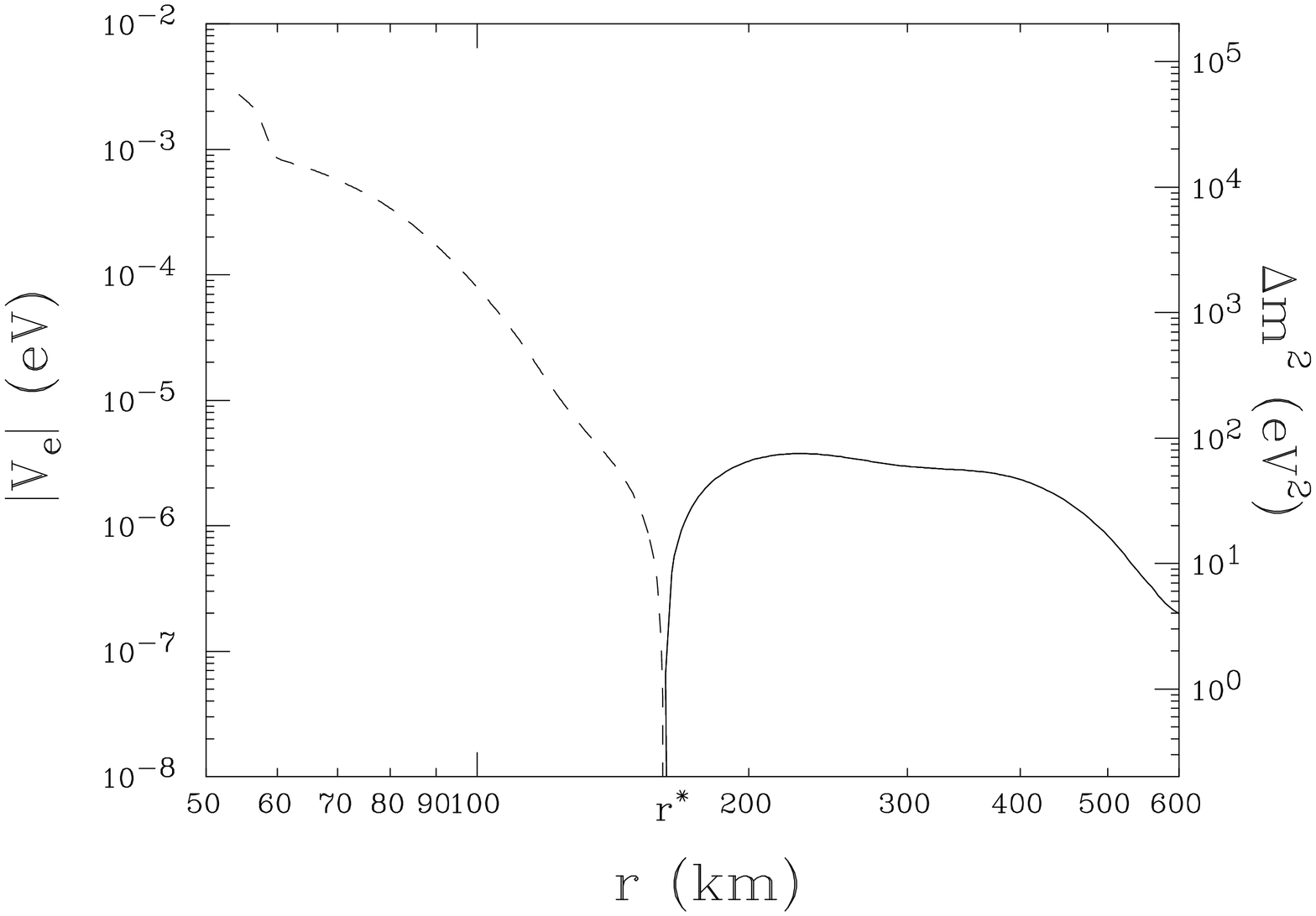,height=6.5cm,width=8.5cm,angle=90}}}
\vglue -0.5cm
\noindent
\caption{Typical $|V_e|$ profile versus the radial distance $r$ at $t<1$ s
after the core bounce. The solid and dashed lines correspond to
positive and negative potential, respectively, and the position where
$V_e=0$ is denoted by $r^*$.  The $\Delta m^2$ values for which a
$E=10$ MeV neutrino undergoes resonant conversion, for the $|V_e|$
value on the left ordinate is indicated in the right ordinate. This is
relevant for the shock revival considerations.}  
\label{potential1}
\eef
\vglue -0.5cm
\bef
\centerline{\protect\hbox{
\psfig{file=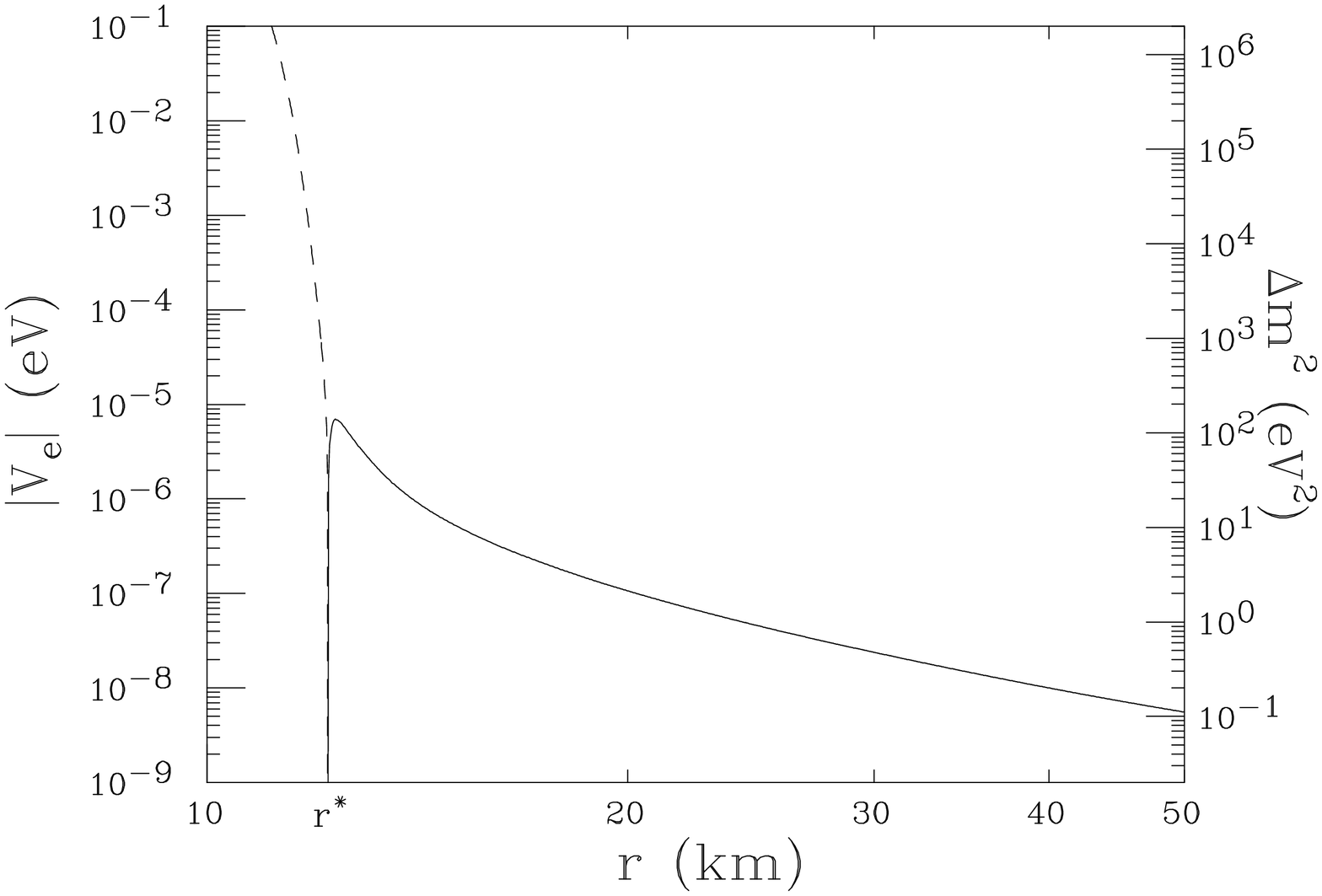,height=6.5cm,width=8.5cm,angle=90}}}
\vglue -0.8cm
\noindent
\caption{Same as Fig. 1 for  $t>1$ s. This is relevant for the
$r$-process and $\bar{\nu}_e$ signal considerations.
}
\eef
%%%%%%%%%%%%%%%%%%%%%%%%%%%%%%%%%%%%%%%%%%%%%%%%%%%%%%%%%%%%%
In Figs. 1 (earlier epoch) and 2 (later epoch) we reproduce 
the typical profiles of corresponding $|V_e|$ in 
the region of interest, taken from ref. \cite{sterilenprv}. 
They are obtained from the $Y_e$ and $\rho$ profiles taken from
Wilson's supernova model outside the neutrinosphere, at two different
times, $t< 1$ s and $t> 1$ s post-bounce.  There is a
point where $Y_e$ takes the value $1/3$ (i.e. $V_e=0$), as indicated
by $r^*$. This position corresponds to $r^* \approx 160 $ km and 12 km
for the earlier and later epochs, respectively.  Clearly, the
effective potential $V_e$ changes its sign from negative to positive
at the point $r^*$.

The resonance condition in \eq{rc} provides the $\Delta m^2$ value for
which neutrinos with some given energy can experience the resonance
for a certain value of the potential (or equivalently, at some
position $r$). For convenience, in the right ordinate of 
the Figs. 1 and 2 
we have also indicated the corresponding $\Delta m^2$ values 
for which neutrinos under go resonance, for typical neutrino 
energy $E= 10$ MeV.
We see that for $\Delta m^2 \gsim 10^{2}$ eV$^2$ only $\bar{\nu}_e \to
\bar{\nu}_s$ conversions can occur, and this happens in the region
where $Y_e \leq 1/3$. 
On the other hand for $\Delta m^2 \lsim 10^{2}$ eV$^2$ three
resonances may occur. The $\bar\nu_e$'s are first converted, say at
$r_1 < r^*$, then there are two resonance points at $r_2\  (\approx r_1)$ 
and $r_3$, where $r_3> r_2 >r^*$ (i.e. in the region where $Y_e > 1/3$), 
for the $\nu_e \to \nu_s$ channel. A schematic level crossing diagram for
${\nu}_e-{\nu}_s$ and $\bar{\nu}_e-\bar{\nu}_s$ system may be seen
explicitly in Fig. 3 of ref.  \cite{sterilenprv}. Note that the
$\nu_s$'s originated from the first $\nu_e$ conversion (at $r_2$) can
be re-converted into $\nu_e$'s at the second resonance (at $r_3$).

In what follows we will employ the simple Landau-Zener approximation
\cite{Landau,HPD} in order to estimate the survival probability after
the neutrinos cross the resonance. Under this approximation, the
$\nu_e$ (or $\bar\nu_e$) survival probability is given by,
\begin{eqnarray}
\label{LZ}
 P & = & 
\exp\Biggl(-\dfrac{\pi^2}{2}\dfrac{\Delta r}
{L_{m}^{\rm res}} \Biggr) \nonumber \\
          & \approx & \exp\left[
-1.1 \times  10^{-6} %\times (\mu_\nu B)^2
\left(\dfrac{\mu B }{\mu_B \mbox{G}} \right)^2_{res}
\left(
\dfrac{\mbox{d}V_e}{\mbox{d}r}  \times 
\dfrac{\mbox{km}}{\mbox{eV}}
\right)^{-1}_{res}
                 \right] \ , %\nonumber \\
% \Delta r & = & \dfrac{\mbox{d}r}{\mbox{d}V_e}\mbox{d}V_e ~~ , 
\end{eqnarray}
where $L_{m}^{\rm res}$ is the neutrino oscillation length at 
resonance.  
Notice that for $\Delta r /L_{m}^{\rm res}>1$ the resonant neutrino
conversion will be adiabatic \cite{MSW}.

As we discussed in ref. \cite{sterilenprv}, depending on the $\Delta
m^2$ values, the $\nu_e-\nu_s$ system may encounter two resonances.
In order to take this into account we compute the $\nu_e$ survival
probability after the second resonance as follows,
\begin{equation}
\label{prob}
P(\nu_e \to  \nu_e ) = P(r_2) P(r_3) + [1-P(r_2)] [1-P(r_3)],
\end{equation}
where $P(r_2)$ and $P(r_3)$ are the survival probabilities calculated
according to the \eq{LZ} at the first and second resonance positions
$r_2$ and $r_3$, respectively.

In this work we will neglect the spin precession due to the
galactic magnetic field discussed in ref. \cite{snbound}, since for
the relevant parameters of interest to us here the condition $\mu_\nu
B_G \ll \Delta m^2/E$, where $B_G \sim 10^{-6}$ G is the galactic
magnetic field, is always satisfied so that the spin precession in the
galactic field is strongly suppressed.

\section{Implications for Supernova Physics}

Following ref. \cite{sterilenprv} we now consider the possible impact
of active-sterile neutrino conversions on different aspects of
supernova astrophysics. We analyse processes taking place at the early
epochs as well as at the cooling stage, in order to get a feeling for
their sensitivity to the underlying neutrino mass-square difference,
magnetic moments and magnetic fields. In the following we consider
only epochs after the core bounce and the neutrino evolution in
regions outside the neutrino-spheres, where resonant conversions take
place.
\subsection{Shock Revival}
Here we are considering the earlier epoch $t< 1$ second after core
bounce.  In the delayed explosion scenario \cite{CW,Wilson85} the
neutrino energy deposition, occurring between the neutrino sphere and
the site where the shock is stalled (about 500 km or so from the core)
can re-start the shock and power the explosion. The disappearance of
either $\nu_e$ or $\bar{\nu}_e$ due to the resonant spin precession
into sterile states would reduce the energy deposition rate.
The neutrino energy deposition rate $R$ at the stalled shock is
defined as follows,
\begin{eqnarray}
\label{ratio}
 R & = &\frac{ Y^{\prime}_n   \dot{E}^\prime_{\nu_e n}(t) + 
Y^\prime_p   \dot{E}^\prime_{\bar\nu_e p}(t) } 
{ Y_n   \dot{E}_{\nu_e n}(t) + 
Y_p   \dot{E}_{\bar\nu_e p}(t) } ~~, \\
   \dot{E}_{\nu N}(t) & \sim &
 \int E_\nu \sigma_{\nu N} \phi^0 (E_\nu)\mbox{d}E_\nu ~~, ~~~~~
\dot{E}^\prime_{\nu N}(t)  \sim 
 \int E_\nu \sigma_{\nu N} P(E_\nu)
 \phi^0(E_\nu)\mbox{d}E_\nu ~.  
\label{ratio2}
\end{eqnarray}
where the primed quantities $Y^\prime_p$ and $Y^\prime_n$ stand for
the proton and neutron abundances calculated in the presence of
active-sterile neutrino conversions. The un-primed ones correspond to
the case where no conversion occurs. As an approximation, we have
assumed, in \eq{ratio2} that the neutrino energy spectra
$\phi^0(E_\nu)$ are Fermi-Dirac with zero chemical potential but with
different characteristic temperatures $T_\nu$ which depend on the
neutrino flavour.  
As in our previous analysis \cite{sterilenprv} we choose 
$T_\nu$ for $\nu_e$ and $\bar{\nu}_e$ such that the typically 
predicted average energies, $\vev{E_{\nu_e}} = 11$ MeV and 
$\vev{E_{\bar{\nu}_e}} = 16$ MeV are obtained in the absence of 
any conversion. 
We assume a magnetic field profile $B(r) = B_0
(r_0/r)^n$ where $B_0 = 10^{16}$ G, $n=2$, $r_0 = 100 $ km and $r$ is
the distance from the center of the star. We have also studied the
case where $n=3$ and obtained the similar results.

\bef
\centerline{\protect\hbox{
\psfig{file=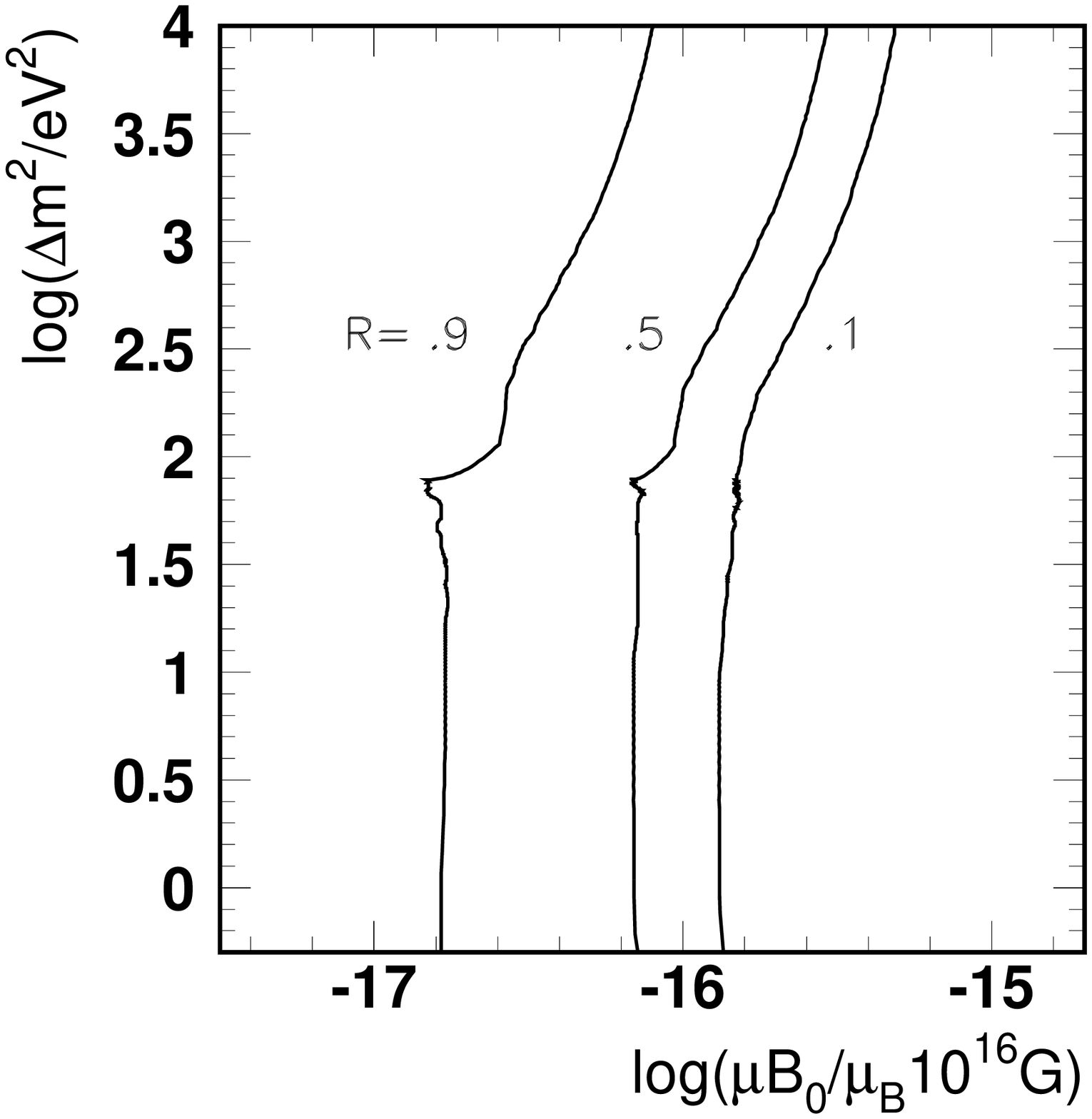,height=6.0cm,width=6.5cm}
}}
\noindent
\caption{Contour plot of the reheating rate $R$ in the $\Delta m^2 -
\mu B_0$ plane. The rate is normalized to 1 in the absence of any kind
of neutrino conversion. Here $\mu_B$ denotes the Bohr magneton.}
\label{shock}
\eef
In \fig{shock} we plot the iso-contour of $R$ in terms of
($\Delta m^2, \mu B_0$).
We show the values of $\mu_\nu B_0$ in units of $\mu_B\cdot 10^{16}$
Gauss, where $10^{16}$ G (at $r = 100$ km) might be the maximally
conceivable value \cite{TD}. 
This is taken as a reference value in order to illustrate the maximal 
sensitivity to the neutrino magnetic moment (though such a value of 
magnetic field may affect the supernova dynamics and self-consistent
calculations would be required for its justification \cite{janka}).
Thus we can say that if the neutrino re-heating is essential for
successful supernova explosion the parameter region right to the
curve, say $R=0.5$, is disfavoured.

\subsection{Implications for the detection of SN1987A $\bar\nu_e$ signal}
 
We are now in the later epoch $t >1$ second after core bounce, in the
so-called Kelvin-Helmholtz cooling phase.  The observed $\bar\nu_e$
events from the supernova SN1987A in Kamiokande II \cite{KA} and IMB
\cite{IMB} detectors, 11 and 8 events, respectively, are in good
agreement with the theoretical expectations.  Therefore, any
significant conversion of $\bar\nu_e$'s into a sterile neutrino
species would be in conflict with this evidence. We consider the
effect of active-sterile neutrino conversions both on the $\bar\nu_e$
signal in order to analyse the possible restrictions on neutrino
parameters, the neutrino magnetic moment and the mass-squared
difference.

We plot in \fig{SN87} three contours of the $\bar\nu_e$ survival
probability $\vev{P}$, which is averaged over the Fermi-Dirac energy
distribution, for the $\bar \nu_e \to \bar \nu_s$ conversion, in the
($\Delta m^2, \mu B_0$) plane.
In the plot the left, the middle and the right lines correspond to 
$\vev{P}$=0.9, 0.5 and 0.1, respectively. 
If we assume that the successful observation of the SN1987A signal
implies that at least 50 \% of the expected $\bar\nu_e$ signal has
been detected, one can conclude that all the portion right to the
curve $P=0.5$ is ruled out.
We note that due to the fact that the potential is much steeper in the
later epoch than in the earlier epoch, the bound we obtained in Fig. 4
is much weaker than the sensitivity we have displayed in Fig. 3. 

%%%%%%%%%%%%%%%%%%%%%%%%%%%%%%%%%%%%%%%%%%%%%%%%%%%%%%%%%%%%%%%%%%%
\bef[t]
%\vglue -1cm
\centerline{\protect\hbox{
\psfig{file=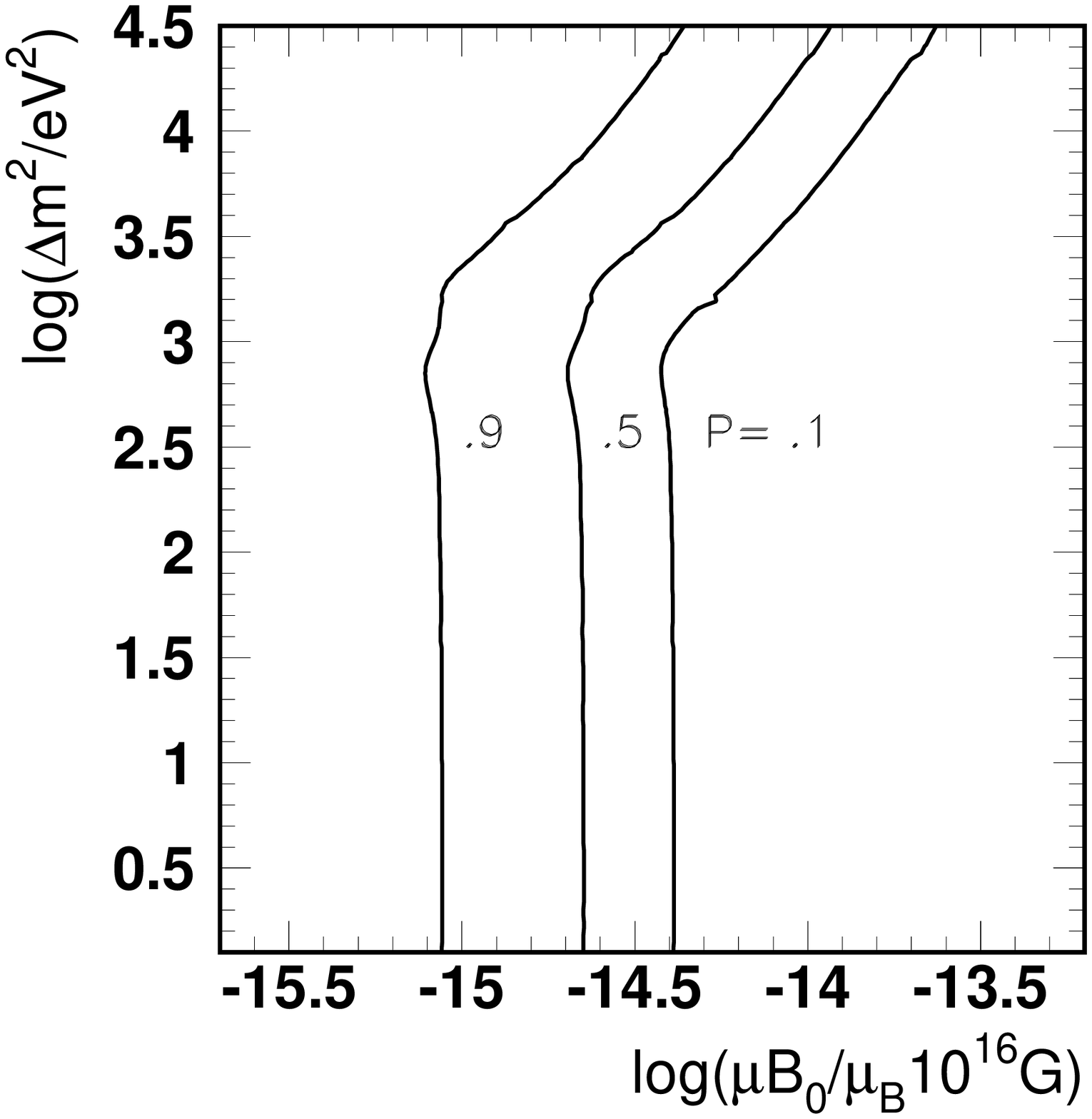,height=6.0cm,width=6.5cm}
}}
\noindent
\caption{Contour plots of the survival probability $P$ (figures at the
curve) for the $\bar\nu_e \to \bar\nu_s$ conversion at $t>1$ s after
the bounce.
The region to the right of the curve corresponding to $P=0.5$ can be
excluded by the observation of the SN1987A $\bar\nu_e$ signal. We have
assumed the magnetic profile to be $B(r) = B_0 (r_0/r)^n$ where $r_0$
= 10 km and $n=2$.}
\label{SN87} 
\eef
%%%%%%%%%%%%%%%%%%%%%%%%%%%%%%%%%%%%%%%%%%%%%%%%%%%%%%%%%%%%%%%%%

\subsection{Implications for $r$-process Nucleosynthesis}

Here, we consider the implications of resonant neutrino conversions
for the supernova nucleosynthesis of heavy elements
\cite{rprocess,sterilenprv}.  It has been proposed that the supernova
could be the most promising site for heavy elements (mass number
$A>70$) nucleosynthesis, the so-called $r$-process \cite{Woosley}, in
which neutrinos play very important r\^ole \cite{qian}.  A necessary
condition required for the $r$-process is $Y_e<0.5$ in the
nucleosynthesis region.  The $Y_e$ value in the $r$-process site is
mainly determined by the following neutrino absorption reactions,
\begin{eqnarray}
\label{nu-n}
\nu_e+n & \to & p+e^, \\
\label{nu-p}
\bar\nu_e+p & \to & n +e^+. 
\end{eqnarray}
Therefore, in the nucleosynthesis region we can write $Y_e$ as follows
\cite{rprocess}, 
\begin{equation}
\label{ye_new}
Y_e \approx {1\over 1+P_{\bar\nu_e}{\VEV{E_{\bar\nu_e}}}/
P_{\nu_e}\VEV{E_{\nu_e}}},
\end{equation}
{}From this expression, we notice that the $\bar\nu_e \to \bar\nu_s$
conversion leads to an increase of $Y_e$, whereas the $\nu_e \to
\nu_s$ conversion acts in the opposite way.
As we described in detail in ref. \cite{sterilenprv}, depending on the
$\Delta m^2$ range, one channel dominates over the other one, and
$Y_e$ can be increased or decreased.
Using again the Fermi-Dirac energy distribution to average the
neutrino absorption rates, we have calculated the electron abundance
$Y_e$ at the site where heavy elements nucleosynthesis is expected to
take place as functions of $(\Delta m^2, \mu B_0)$. 

We present our result in Fig. 5. 
For a successful $r$-process, the region above $Y_e > 0.5$ is ruled
out.  On the other hand we find that the supernova nucleosynthesis
could be enhanced in the region enclosed by the contour $Y_e=0.4$,
similar to the results obtained in ref. \cite{sterilenprv}. 
This region is delimited by
$\Delta m^2 \lsim 10^2$ eV$^2$ and $10^{-16} \lsim \mu_\nu \lsim
10^{-15} $ in units of $\mu_B$ for $B_0 = 10^{16}$ G.
Inside the contour of $Y_e=0.33$, $Y_e$ might get stabilised to $~$
1/3 due to some feedback effect which we discussed in detail in
ref. \cite{sterilenprv}.  We see from the plot that the most promising
mass range for the neutrino hot dark matter scenario \cite{cobe2},
$\Delta m^2 \lsim 10$ eV$^2$, lies inside that where the $r$-process
nucleosynthesis is enhanced. Moreover, it is neither in conflict with
the re-heating process (see Fig. 3), 
for $\mu_\nu B_0 \lsim (10^{-15} \mu_B) \cdot (10^{16}$ G),  
nor with SN1987A observations (see Fig. 4).

%%%%%%%%%%%%%%%%%%%%%%%%%%%%%%%%%%%%%%%%%%%%%%%%%%%%%%%%%
\bef[t]
\centerline{\protect\hbox{
\psfig{file=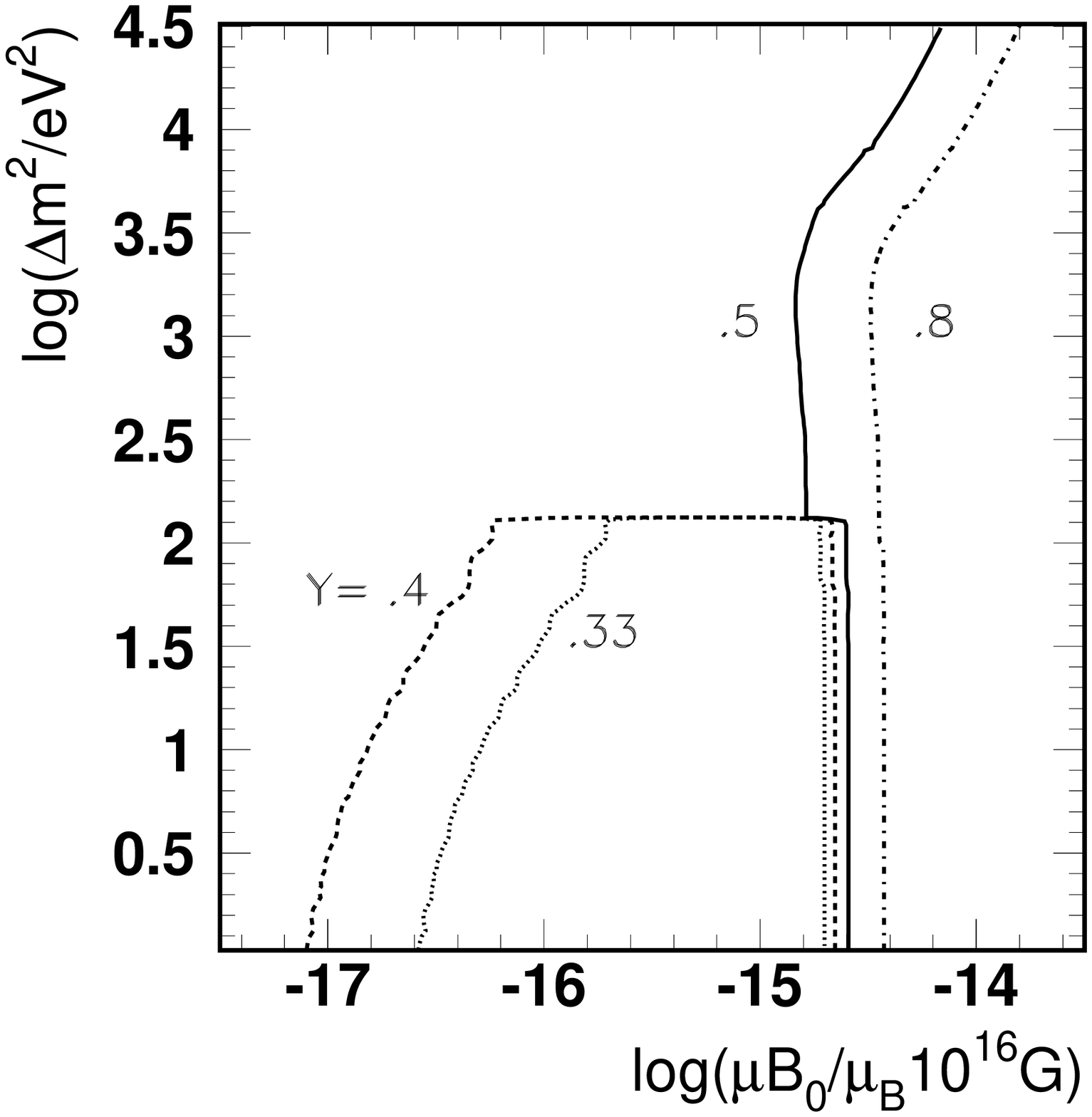,height=6.0cm,width=6.5cm}
}}
\noindent
\caption{
Contour plots for the electron fraction $Y_e$ (figures at the curves)
taking into account $\bar\nu_e \to \bar\nu_s$ and 
$\nu_e \to \nu_s$ conversions at $t>1$ s after the bounce. 
The region to the right of the solid line labelled 0.5 is ruled out by 
the condition $Y_e<0.5$ necessary for $r$-process nucleosynthesis to occur. 
For the parameter region inside the $Y_e= 0.4$ dotted contour $r$-process 
nucleosynthesis can be enhanced.}
\eef
%%%%%%%%%%%%%%%%%%%%%%%%%%%%%%%%%%%%%%%%%%%%%%%%%%%%%%%%

\section{Conclusions and Discussions}

In this paper we have considered the effect on supernova physics of
having RSFP conversions induced by non-zero active-sterile TMM.  We
have investigated the sensitivity of the shock revival and $r$-process
arguments, as well as estimated the restrictions from the observed
SN1987A $\bar{\nu}_e$ detection rates. We have analysed the effect of 
RSFP conversions involving \ne to \ns or \bne to \bns in the region above
the hot proto-neutron star in type II supernovae, assuming the simple
magnetic field profile, $B(r) = B_0 (r_0/r)^n$ $(n = 2,3)$ where $r_0$
is taken to be 100 and 10 km for earlier and later epoch,
respectively.
If the mass of the \ns is in the cosmologically interesting range and
the product of the TMM by the magnetic field just above the
neutrinosphere, $\mu_\nu B_0$, is larger than $\sim (10^{-15} \mu_B)
\times (10^{16}$ G), then a significant fraction of $\nu_e$ and
$\bar{\nu}_e$ would be converted into $\nu_s$ and $\bar{\nu}_s$,
respectively, in the region outside neutrinosphere.
Such conversion could lead to the depletion of $\nu_e$ and
$\bar{\nu}_e$ fluxes, resulting in a suppression of the expected
$\bar{\nu}_e$ signal in underground terrestrial detectors, 
in contradiction with the successful observation of the SN1987A 
$\bar{\nu}_e$ signal in the IMB and Kamiokande detectors. 
Hence, on this basis we can constrain the neutrino mass and 
neutrino magnetic moment by requiring that the total $\bar{\nu}_e$
flux during the thermal neutrino emission epoch should not be
significantly depleted by $\bar{\nu}_e \to \bar{\nu}_s$ conversion, 
and can rule out the range $\mu_\nu B_0 \gsim
(10^{-15}\mu_B) \times (10^{16}$ G).

The TMM-induced \ne - \ns RSFP would also suppress
the neutrino re-heating behind the stalled shock.  We have found that
for $\mu_\nu B_0 \gsim (10^{-16} \mu_B) \times (10^{16}$ G), the
energy deposition by ${\nu}_e$ and $\bar{\nu}_e$ absorption reactions
during the shock re-heating epoch ($t<1$ s after the bounce) could be
significantly decreased.
{}From the $r$-process argument, for the parameter range $\Delta m^2
\gsim 100\ \mbox{eV}^2$, where the $\bar{\nu}_e \to \bar{\nu_s}$
conversion is dominant, $Y_e$ at the nucleosynthesis site could become
larger than 0.5 and hence $r$-process would be forbidden, leading to a
disfavoured range, $\mu_\nu B_0 \gsim (10^{-15}\mu_B) \times (10^{16}$
G).
On the other hand, for $\Delta m^2 \lsim 100 \mbox{ eV}^2$,
$r$-process nucleosynthesis could be enhanced due to the decrease of
$Y_e$ down to the minimum value 1/3 due to the efficient $\nu_e \to
\nu_s$ conversion if the parameters are in the region
%$10^{-16} \lsim \mu_\nu \lsim 10^{-15} $ in units of $\mu_B$. 
$ (10^{-16}\mu_B) \times (10^{16} \mbox{G})
\lsim \mu_\nu B_0 \lsim (10^{-15} \mu_B) \times  (10^{16} $G). 

Throughout the above discussion we have neglected, for simplicity, the
production of the sterile state due to the neutrino helicity-flipping
processes in the core, which was discussed in refs. \cite{snbound,Ayala}.  
This requires $\mu_\nu \lsim 10^{-12} \mu_B$.
Let us finally discuss briefly the case $\mu_\nu$ is larger than 
$10^{-12} \mu_B$ (but smaller than the laboratory limit (1)).  
As discussed in refs.~\cite{snbound,Ayala}, in this case the sterile states 
can be copiously produced by the helicity-flipping scattering processes 
in the SN core and they can escape freely to the outer region with higher
energy. 

For the earlier epoch, relevant for shock reheating, if the parameters
($\Delta m^2, \mu B_0$) are in the region where the RSFP is efficient,
the reheating rate would be increased ~\cite{voloshin} due to the
re-conversion of $\bar{\nu}_s \to \bar{\nu}_e$ and/or $\nu_s \to \nu_e$
since the $\nu_e$ and $\bar{\nu}_e$ would have higher energy compared
to the standard case.  Therefore, we would not get any disfavoured
region but a positive effect from this conversion.
However, for the discussion on the $\bar{\nu}_e$ signal we still would
get constraints due to a different reason \cite{SSB}.  In this case we
would not have the reduction of the $\bar{\nu}_e$ signal but instead
we would get an increase in the expected number of events at
underground detectors. This enhanced $\bar{\nu}_e$ signal would follow
from the increase in the average energy due to the $\bar{\nu}_s \to
\bar{\nu}_e$ re-conversion. This would allow us to disfavour a similar
range of parameters as in Fig. 4. 
Finally from the $r$-process argument, we would have a different
conclusion from what we obtained in Sec. III C. 
Roughly speaking, the allowed region ($Y_e < 0.5$) in Fig. 5 would 
now become disfavoured region whereas the disfavoured region 
($Y_e > 0.5$) in the same plot would now become the allowed one.
The reason is the following. 
For example, if the parameters are in the region where $Y_e < 0.4$ in
Fig. 5, only the resonant conversion at $r_3$ (defined in Sec. II) is
adiabatic due to the fast variation at $r_2$ but the slower variation
of the potential at $r_3$ (see Fig. 2). In the anti-neutrino case, the
$\bar{\nu}_s \to \bar{\nu}_e $ conversion at $r_1$ can be neglected,
as one can see from Fig. 4, for magnetic moments smaller than
$10^{-15}$ $\mu_B$ or so.
This implies that, instead of the net effect of the $\nu_e \to \nu_s$
conversion expected in the case we considered in Sec. III.C, now we
expect efficient $\nu_s \to \nu_e$ conversion, which makes the average
energy of the resulting $\nu_e$ higher than in the no-conversion
case. This has the result of driving the $Y_e$ value larger than 0.5
(see eq. (\ref{ye_new})). On the other hand one can expect the
opposite behaviour for the region where $Y_e > 0.5$ (and larger
magnetic moment) the conversions at all the points $r_i$ are expected
to be adiabatic and therefore important. Thus, roughly speaking, the
net effect is the reduction of $\nu_e$ flux, when neutrinos reach the
position where the r-process is occurring. The net effect is that the
$\nu_e$ flux is decreased and the $\bar{\nu}_e$ is increased. This way
we expect that the value of $Y_e$ will become smaller than in the
no-conversion case. Thus (assuming that $\nu_s$ and $\bar{\nu}_s$ are
copiously produced in the core) the region with $Y_e > 0.5$ would now
become allowed by the r-process criterion.

\vspace{0.7cm}

\centerline{\bf Acknowledgement}

We thank H.-T. Janka for correspondence on the magnetic field of the
proto-neutron star. This work has been supported by DGICYT under Grant
N. PB95-1077, by a joint CICYT-INFN grant, and by the TMR network
ERBFMRXCT960090 of the European Union.  H. N. has been supported by a
DGICYT fellowship at Univ. de Valencia when major part of this work
was done, and by a postdoctral fellowship from Funda\c{c}\~ao de
Amparo \`a Pesquisa do Estado de S\~ao Paulo (FAPESP).  R. T. has been
supported by a fellowship from Generalitat Valenciana.

%\noindent
%\newpage

\end{document}